\documentclass[
]{ceurart}

\usepackage{amssymb}
\usepackage{url}

\usepackage{listings}
\usepackage[symbols,nogroupskip,sort=none]{glossaries-extra}
\usepackage{stmaryrd}
\usepackage{amsmath, nccmath}
\begin{document}

\copyrightyear{2022}
\copyrightclause{Copyright for this paper by its authors.
  Use permitted under Creative Commons License Attribution 4.0
  International (CC BY 4.0).}

\conference{ITASEC 2024: The Italian Conference on CyberSecurity, April 08-12, 2024, Salerno, Italy}

\title{Building Call Graph of WebAssembly Programs via Abstract Semantics}

\author[1]{Mattia Paccamiccio}[%
orcid=0000-0003-2096-0191,
email=mattia.paccamiccio@unicam.it,
]
\cormark[1]
\address[1]{Università di Camerino, Via Andrea d'Accorso 16, 62032, Camerino, Italy}

\author[2]{Franco Raimondi}[%
orcid=0000-0002-9508-7713,
email=f.raimondi@gssi.it,
]
\address[2]{Gran Sasso Science Institute, Viale Francesco Crispi 7, 67100, L'Aquila, Italy}

\author[1]{Michele Loreti}[%
orcid=0000-0003-3061-863X,
email=michele.loreti@unicam.it,
]

\cortext[1]{Corresponding author.}

\begin{abstract}
WebAssembly is a binary format for code that is gaining popularity thanks to its focus on portability and performance. 
Currently, the most common use case for WebAssembly is execution in a browser. It is also being increasingly adopted as a stand-alone application due to its portability. 
The binary format of WebAssembly, however, makes it prone to being used as a vehicle for malicious software. For instance, one could embed a cryptocurrency miner in code executed by a browser. 
As a result, there is substantial interest in developing tools for WebAssembly security verification, information flow control, and, more generally, 
for verifying behavioral properties such as correct API usage. In this document, we address the issue of building call graphs for WebAssembly code. 
This is important because having or computing a call graph is a prerequisite for most inter-procedural verification tasks. In this paper, we propose a formal 
solution based on the theory of Abstract Interpretation.
We compare our approach to the state-of-the-art by predicting how it would perform against a set of specifically crafted benchmark programs.
\end{abstract}

\begin{keywords}
  WebAssembly \sep
  Call Graph \sep
  Abstract Interpretation \sep
  Semantics \sep
  Program Analysis
\end{keywords}

\maketitle

\newcommand{\tool}{\textit{wassilly}~}

\glsxtrnewsymbol[description={WebAssembly module}]{mod}{\ensuremath{M}}
\glsxtrnewsymbol[description={Program (instruction, function)}]{P}{\ensuremath{P}}
\glsxtrnewsymbol[description={Environment}]{Env}{\ensuremath{\epsilon}}
\glsxtrnewsymbol[description={Abstract environment}]{aEnv}{\ensuremath{\widehat{\epsilon}}}
\glsxtrnewsymbol[description={Abstract memory}]{aMem}{\ensuremath{\widehat{\sigma}}}
\glsxtrnewsymbol[description={Memory}]{Mem}{\ensuremath{\sigma}}
\glsxtrnewsymbol[description={Continuation}]{chi}{\ensuremath{\chi}}
\glsxtrnewsymbol[description={Transformation}]{tau}{\ensuremath{\tau}}
\glsxtrnewsymbol[description={Interpreter}]{i}{\ensuremath{\mathcal{I}}}
\glsxtrnewsymbol[description={Label}]{l}{\ensuremath{l}}
\glsxtrnewsymbol[description={Map $l\,\xrightarrow\,(\tau,\chi)$}]{rho}{\ensuremath{\rho}}
\glsxtrnewsymbol[description={Timestamp}]{phi}{\ensuremath{\phi}}
\glsxtrnewsymbol[description={Context}]{k}{\ensuremath{\kappa}}
\glsxtrnewsymbol[description={Abstract context}]{aK}{\ensuremath{\widehat{\kappa}}}
\glsxtrnewsymbol[description={Concretization}]{gamma}{\ensuremath{\widehat{\gamma}}}
\glsxtrnewsymbol[description={Union}]{sqcup}{\ensuremath{\widehat{\sqcup}}}
\glsxtrnewsymbol[description={Widening}]{widen}{\ensuremath{\widehat{\nabla}}}

\section{Introduction}\label{sec:intro}

WebAssembly \emph{``is a binary instruction format for a stack-based virtual machine, designed as a 
portable compilation target for programming languages"}\footnote{\url{https://webassembly.org/}}. 
WebAssembly aims to be portable and efficient: it is actively supported by most web browsers and it 
is employed for rendering streamed videos on devices such as smart televisions, set-top boxes, and 
other embedded devices~\cite{alexene_wasm,schroeder_disneyplus}.

WebAssembly is primarily used in conjunction with JavaScript to optimize tasks that are typically 
resource-intensive and would benefit from more optimized, close-to-native performance~\cite{alexene_wasm}, 
such as rendering images, streaming videos and generally heavy computational tasks for which 
interpreted languages might be too slow. 

In parallel with its adoption, researchers have investigated the issue of verifying WebAssembly code 
directly~\cite{toughcall}. From a security point of view, this is motivated by the fact that WebAssembly, while 
executing in a relatively safe sandboxed environment, is still the source of security concerns 
(information leakage, tampering, policy violation, unauthorized use of computational power, etc.).

In terms of formal analysis, a call graph is a prerequisite for verifying several properties of 
interest, such as information flow analysis, call-ordering requirements, safety, and liveness. 
We define a call graph as a directed graph that encodes the possible calling relationships between 
functions or methods in a program, represented as nodes, and edges between two nodes (i.e. $f$, $g$) 
representing that $f$ calls $g$. Constructing an \emph{exact} call graph is an undecidable problem. 
The most adopted approach is soundly approximating all the possible calls, implying that an edge 
between $f$ and $g$, in a sound approximation, translates to $f$ \emph{may} call $g$. The precision 
of a call graph depends on the inference rules used, which could include or ignore information 
about execution paths, context, and so on.

\subsection{Contribution}\label{subsec:contribution}

We investigate the problem of building a precise call graph for WebAssembly. The simple execution 
model of WebAssembly, briefly described in Section~\ref{sec:background}, implies the need to track 
values on top of the stack to resolve call sites relying on some form of indirection. 
Other sound approaches can be improved in precision ~\cite{stievenart2021wassail,sturdy_wasm_ai2023}, while precise approaches are unsound 
(MetaDCE\footnote{\url{https://github.com/WebAssembly/binaryen}}, 
WAVM\footnote{\url{https://wavm.github.io/}}, Manticore~\cite{mossberg2019manticore}).  

We propose an approach based on the theory of Abstract interpretation, aiming to improve precision 
while retaining soundness.

We lay out the paper as follows: we introduce 
background details about WebAssembly and Abstract interpretation in Section~\ref{sec:background}; 
we describe our formal approach in Section~\ref{sec:exper};  we present related work in Section~\ref{sec:related} and we conclude in Section~\ref{sec:conclusion}.

\section{Background}\label{sec:background}

In this section, we start by giving an overview of WebAssembly. We then introduce basic notions of 
call graphs, what soundness entails when analyzing WebAssembly modules, and Abstract interpretation. 

\subsection{Basics of WebAssembly}\label{subs:wasm_basics}

WebAssembly is a simple, typed, assembly-like language whose execution model is based on a stack
machine, implementing a set of low-level instructions, less than 200 in total. It is designed to be 
a universal compilation target, mostly for the LLVM family of programming languages. The resulting 
binary is then run on an ad-hoc Virtual Machine (VM). The most common example of WebAssembly 
Virtual Machine as of today is a sandboxed environment in a browser, but stand-alone runtime 
environments for resource-constrained devices are available\footnote{
  \url{https://github.com/bytecodealliance/wasm-micro-runtime}
}. Overall, the paradigm is 
very similar to the one that Java adopts: source code is compiled into bytecode and then executed 
by a virtual machine. Such a paradigm allows the code to be executed, in theory, on every 
platform, as long as a VM is available.

WebAssembly manages intra-procedural control flow predictably by using labeled jump-like 
instructions\footnote{
  \url{https://webassembly.github.io/spec/core/syntax/instructions.html\#syntax-instr-control}
}~\cite{stievenart2020compositional}. This allows for the production of a 
correct control flow graph of a function by simply analyzing the syntax, whereas in more traditional 
machine code, such as x86 Assembly, control flow graph recovery is an undecidable 
problem~\cite{nguyen2013binarycfg,montoya2020ddisasm}.

Notice also that, in WebAssembly, function calls are performed by either the \verb|call| 
operator or the \verb|call_indirect| operator: the latter can be compared to calling a function 
pointer and, as we will see below, is the only source of lack of precision in call graph 
construction. In WebAssembly function ``names" actually are indices, as such, both of call 
instructions work on numerical indices, either passed directly to the instruction as in 
\verb|call|, or via the stack in \verb|call_indirect| and referring to an index in a table. 

\subsection{Call graphs}

Formally, the call graph of a program is a directed graph $G=(V, E)$ in which the set of nodes $V$ 
is the set of functions in the program, and there exists an edge $(f,g) \in E \subseteq V \times V$ if function $f$ 
calls function $g$ in the program. Apart from the verification of properties such as call ordering 
constraints, the availability of a call graph is a requirement in many other inter-procedural 
analyses that need to track the flow of data in a program.

The construction of an \emph{exact} call graph is an undecidable problem, and call graphs are 
usually computed by over-approximating the possible calls (i.e., the set $E$ contains edges 
containing all the actual calls, and additional spurious edges). Dynamic features such as 
function pointers in C/C++ further complicate the issues of computing a call graph. WebAssembly 
implements function pointers using tables containing references to functions.

In object-oriented programming languages, the main approach to static call graph construction is 
Class Hierarchy Analysis~\cite{dean1995optimization} and its improvements. The main 
idea of Class Hierarchy Analysis is to employ the inheritance hierarchy to refine the set of methods that could be 
invoked at a specific program location. WebAssembly does not have a class hierarchy, so this 
method is not directly applicable. Instead, tools like \emph{Wassail}~\cite{stievenart2021wassail} 
and \emph{Wasmati}~\cite{wasmati} over-approximate calls by looking at the type of the function 
appearing in a \verb+call_indirect+ instruction (i.e., the content of the stack is ignored).
A more granular call graph can only be constructed  by tracking the values that are on top of 
the stack when \verb+call_indirect+ is used.

\subsection{Threats to soundness}\label{subs:threats_validity}
WebAssembly modules are designed with host-module interoperability in mind. One of the possible interactions they can have includes the host performing mutations of the running instance of the WebAssembly module. To be more specific, instances of shared 
memories and tables can be mutated, entailing that with no information available to the analysis 
about what the host code does, the analysis result may be unsound. 
For example, our property of interest is the call graph. If the target of a \verb+call_indirect+ 
instruction depends on a value from a memory block mutated by the host code, the analyzer cannot 
infer this behavior. 
As another example, we could consider a table containing function references modified by the host 
at some point during the execution. 
For the sake of this work, we assume that the module under examination is \emph{closed} and 
address this issue in Section \ref*{sec:futurework}.

\subsection{Abstract interpretation}
Abstract interpretation~\cite{cousot_ai} is a theory of sound approximation of the semantics 
of computer programs. It provides a methodology for defining rigorously sound static analyses. 
Its goal is to derive information about a program's semantics without performing an exhausting 
exploration of the program under analysis. We obtain this by making the semantics decidable at the 
cost of precision (Rice's theorem, halting problem) but in a way that the semantics remain sound compared to 
the concrete semantics. If a concrete interpretation is the evaluation of concrete semantics, an 
abstract interpretation is the evaluation of abstract semantics. How precise abstract semantics 
are when compared to concrete semantics is a matter of compromise between computability and 
tractability of the problem. Due to this factor, tailoring the abstract semantics to the program 
properties of interest can be an optimal design choice.
Abstract interpretation sees application in proving software properties, such as the absence of 
critical errors, including detecting division-by-zero,  NULL pointer dereferencing, etc.

Abstract interpretation makes wide use of upper bound operators, denoted by \gls*{sqcup} and used to compute control-flow joins,
and widening, denoted by \gls*{widen} and used in the computation of loops and recursion in a way that ensures termination.
These operators are sound, and they influence the precision of the result of the analysis.

For WebAssembly, to-date the only known abstract interpreter capable of producing a call graph 
is \emph{Sturdy}~\cite{sturdy_wasm_ai2023,sturdy_combinatorfix}. It uses only non-relational 
abstractions, which can affect the precision of the analysis. 
\emph{Wassail}~\cite{stievenart2021wassail}, too, uses notions of abstract interpretation 
to perform taint analysis.
\section{\tool}\label{sec:exper}
\tool (\textit{WebAssembly Ain't So Silly}) is an abstract interpreter that we are actively developing. It is tailored 
to constructing WebAssembly call graphs but can be extended to verify more general properties, 
such as the absence of runtime errors.

We introduce and briefly explain the architecture of the tool via the semantics we construct
and some of the peculiarities we design for this implementation. 
We omit the abstract semantics of particular instructions if they are homomorphic to their 
concrete counterparts.

\subsection{Semantics}
We introduce concrete and abstract denotational semantics for a simplified version of WebAssembly. Semantics is the rigorous definition of the meaning of the syntactic constructs of a programming language.

The form of semantics we use employs an explicit passing of the next function to compute and is known as continuation semantics. We denote continuations with the symbol \gls*{chi}.

In the context of our semantics, we define a memory \gls*{Mem}, holding: a store $se$ comprising of
bindings of values to global variable declarations; an indexed table referencing functions; a reference to the topmost element (in-context) in the call (or context) stack, \gls*{k}.
The global store and in-context value stack and local variables are available for reading and updating via specific accessors embedded in \gls*{Mem}, respectively: $.se$, $.sk$, $.loc$.

We use an environment \gls*{Env} to keep track of local and global variable declarations.

The symbol \gls*{phi} is used as a timestamp to differentiate running function instances.
\gls*{rho} is a map from label statements to a tuple consisting of a continuation \gls*{chi} and a 
transformation \gls*{tau} happening in case branching instructions such as \verb|br| or \verb|br_if| are interpreted with any given label $l$ as the argument.

Our semantics take as inputs: \gls*{rho}, \gls*{chi}, \gls*{Mem}, \gls*{Env}, \gls*{phi}, \gls*{k} and produce a function consisting of the continuation \gls*{chi} taking as input the new values for \gls*{rho}, \gls*{Mem}, \gls*{Env}, \gls*{phi}, \gls*{k}. The ``normal" continuation \gls*{chi} might be overridden with $\gls*{chi}'$ in case of non-linear control flow.

The semantics we defined cover what we deem as the main language concerns, keeping aside \verb*|memory| operations, which behave similarly to global variables, hence why they are excluded from this work. 
Typing is assumed always to be \verb*|i32|, 32-bit integer.

\paragraph{Auxiliary functions and notations}
In order to ease the reading of the semantics we define a set of auxiliary functions. 
These include basic operations on the stack, boolean evaluation, variable allocation, and a timestamp 
to differentiate function instances. Some functions that will appear in the semantics will be self-explanatory (e.g: $filter$, $bind$) and are not included in this table.

\begin{longtable}{l | c}

    $let\ pop\ \alpha\ =\ (\alpha', e)$ & Pop an element (value, label) from the stack. \\ \hline
    
    $let\ push\ \alpha\ \Tilde{\delta}\ =\ \alpha'$ & Push an element (value, label) to the stack. \\ \hline
    
    $let\ pop\_twice\ \gls*{Mem}\ =\ $\\
    \phantom{XX}$let\ \gls*{Mem}',\ v_{right}\ =\ pop\ \gls*{Mem}\ in$ & $pop$ is performed twice.\\
    \phantom{XX}$let\ \gls*{Mem}'',\ v_{left}\ =\ pop\ \gls*{Mem}'\ in$\\
    \phantom{XX}$(\gls*{Mem}'', v_{left}, v_{right})$ & \\ \hline
    
    $let\ pop\_n\ \gls*{Mem}\ n\ = (\gls*{Mem}', \Tilde{v})$ & $pop$ is performed $n$ times. \\ \hline
    
    $let\ intbool\ \gls*{Mem}=\ $ & Concrete int-to-boolean evaluation:\\
    \phantom{XX}$let\ \gls*{Mem}',v=pop\ \gls*{Mem}\ in$ & 0 is $false$, any non-zero value is $true$. \\
    \phantom{XX}$v=0\ ?\ (\gls*{Mem}',false)\ :\ (\gls*{Mem}',true)$ & \\ \hline
    
    $let\ \widehat{intbool}\ \gls*{aMem}=\ $ & Abstract int-to-boolean evaluation:\\
    \phantom{XX}$let\ \gls*{aMem}',\widehat{e}=pop\ \gls*{aMem}\ in\ \gls*{aMem}', \widehat{e}$ & the expression $\widehat{e}$ is evaluated later. \\ \hline
    $let\ tick\ \gls*{phi}\ =\ \gls*{phi}'$ & Timestamp generator. \\ \hline
    
    $let\ local\ i\ =\ local\ location $ & Allocates a local variable. \\ \hline
    
    $let\ global\ i\ =\ global\ location $ & Allocates a global variable. \\ \hline
    
    \caption{Auxiliary functions, briefly explained.}
    \label{tab:auxfuncs}
\end{longtable}

\pagebreak

\subsubsection{Concrete semantics}

In Table \ref{tab:concretesem} we define the concrete semantics of the reduced version of the WebAssembly instruction set we studied. To avoid ambiguities, we provide a brief description of the semantics. 

\begin{longtable}{l | c}
Concrete semantics & Brief description \\
\hline
\endfirsthead
Concrete semantics & Brief description \\
\hline
\endhead
 
  $\llbracket c_{1};c_{2} \rrbracket_{\gls*{rho}\gls*{chi}\gls*{Mem}\gls*{Env}\gls*{phi}\gls*{k}}\ =\ $ 
  $\llbracket c_{1} \rrbracket_{\gls*{rho}}$
  $({\llbracket c_{2}\rrbracket}_{\gls*{chi}})_{\gls*{Mem}\gls*{Env}\gls*{phi}\gls*{k}}$ & Sequence of commands \\ \hline
  $\llbracket unop \rrbracket_{\gls*{rho}\gls*{chi}\gls*{Mem}\gls*{Env}\gls*{phi}\gls*{k}}\ =$ & Class of operations consisting in: \\
  \phantom{XX}$let\ \gls*{Mem}',\ v\ =\ pop\ \gls*{Mem}.sk\ in$ & pop once, perform a computation, \\
  \phantom{XX}$\gls*{chi}(push\ \gls*{Mem}'\ [unop\ v])\gls*{rho}\gls*{Env}\gls*{phi}\gls*{k}$ & push the result to the stack.\\ \hline
  
  $\llbracket binop \rrbracket_{\gls*{rho}\gls*{chi}\gls*{Mem}\gls*{Env}\gls*{phi}\gls*{k}}\ =$ & Class of operations consisting in:\\
  \phantom{XX}$let\ \gls*{Mem}',\ v_{left},\ v_{right}\ =\ pop\_twice\ \gls*{Mem}.sk\ in $ & pop twice, perform a computation, \\
  \phantom{XX}$\gls*{chi}(push\ \gls*{Mem}'\ [binop\ v_{left}\ v_{right}])\gls*{rho}\gls*{Env}\gls*{phi}\gls*{k}$ & push the result to the stack.\\ \hline
  
  $\llbracket local\ i\ t\rrbracket_{\gls*{rho}\gls*{chi}\gls*{Mem}\gls*{Env}\gls*{phi}\gls*{k}}\ = $\\
  \phantom{XX}$let\ \gls*{Env}',newvar\ =\ new\gls*{Env}\ \gls*{Env}\ \gls*{phi}\ (local\ i)\ in$ & Declaration of local variable\\ 
  \phantom{XX}$let\ \gls*{Mem}'\ =\ bind\ \gls*{Mem}\ \gls*{phi}\ 0\ newvar\ in$ & (typing $t$ is ignored)\\
  \phantom{XX}$\gls*{chi}\gls*{Mem}'\gls*{rho}\gls*{Env}'\gls*{phi}\gls*{k}$ \\ \hline
  
  $\llbracket local.get\ i \rrbracket_{\gls*{rho}\gls*{chi}\gls*{Mem}\gls*{Env}\gls*{phi}\gls*{k}}\ =\ $ & Read the value of a local variable, \\
  \phantom{XX}$\gls*{chi}(push\ \gls*{Mem}.sk\ [\gls*{Mem}.loc(i)])\gls*{rho}\gls*{Env}\gls*{phi}\gls*{k}$ &  then push it to the stack \\ \hline
  
  $\llbracket local.set\ i \rrbracket_{\gls*{rho}\gls*{chi}\gls*{Mem}\gls*{Env}\gls*{phi}\gls*{k}}\ =$ \\
  \phantom{XX}$let\ \gls*{Mem}',v\ =\ pop\ \gls*{Mem}.sk\ in$ & Update a local variable \\
  \phantom{XX}$\gls*{chi}(\gls*{Mem}'.loc[i \leftarrow v])\gls*{rho}\gls*{Env}\gls*{phi}\gls*{k}$ \\ \hline
  
  $\llbracket global\ i\ t \rrbracket_{\gls*{rho}\gls*{chi}\gls*{Mem}\gls*{Env}\gls*{phi}\gls*{k}}\ = $\\
  \phantom{XX}$let\ \gls*{Env}',newvar\ =\ new\gls*{Env}\ \gls*{Env}\ \gls*{phi}\ (global\ i)\ in$ & Declaration of global variable\\
  \phantom{XX}$let\ \gls*{Mem}'\ =\ bind\ \gls*{Mem}\ \gls*{phi}\ 0\ newvar\ in$ & (typing $t$ is ignored)\\
  \phantom{XX}$\gls*{chi}\gls*{Mem}'\gls*{rho}\gls*{Env}'\gls*{phi}\gls*{k}$ \\ \hline
  
  $\llbracket global.get\ i \rrbracket_{\gls*{rho}\gls*{chi}\gls*{Mem}\gls*{Env}\gls*{phi}\gls*{k}}\ =\ $ & Read the value of a global variable,\\ 
  \phantom{XX}$\gls*{chi}(push\ \gls*{Mem}.sk\ [\gls*{Mem}.se.glob(i)])\gls*{rho}\gls*{Env}\gls*{phi}\gls*{k}$ & then push it to the stack.\\ \hline
  
  $\llbracket global.set\ i \rrbracket_{\gls*{rho}\gls*{chi}\gls*{Mem}\gls*{Env}\gls*{phi}\gls*{k}}\ = $\\
  \phantom{XX}$let\ \gls*{Mem}',v\ =\ pop\ \gls*{Mem}.sk\ in $ & Update a global variable.\\
  \phantom{XX}$\gls*{chi}(\gls*{Mem}'.se.glob[i \leftarrow v])\gls*{rho}\gls*{Env}\gls*{phi}\gls*{k}$ \\ \hline

  $\llbracket label\ l\ (c) \rrbracket_{\gls*{rho}\gls*{chi}\gls*{Mem}\gls*{Env}\gls*{phi}\gls*{k}}\ = $ & Evaluation of a labeled block,\\
  \phantom{XX}$let\ \gls*{Mem}',\Tilde{v}\ =\ pop\_n\ \gls*{Mem}.sk\ (arity\ l)\ in $ & it can be either Block or Loop. \\
  \phantom{XX}$let\ \gls*{Mem}''\ =\ push\ (push\ \gls*{Mem}'.sk\ [l])\ \Tilde{v}\ in $ & A $fix$ operator is used to \\
  \phantom{XX}$fix(\lambda\Theta.[\![c]\!]_{\gls*{rho} [l \leftarrow \gls*{tau},\Theta]\ \gls*{chi}\gls*{Mem}'''\gls*{Env}\gls*{phi}\gls*{k}})$&ensure termination in Loops.\\ \hline
  
  $\llbracket br\ l \rrbracket_{\gls*{rho}\gls*{chi}\gls*{Mem}\gls*{Env}\gls*{phi}\gls*{k}}\ = 
    let\ \gls*{tau},\gls*{chi}'\ = \gls*{rho}\ l\ in $ & Unconditional branching.\\
  \phantom{XX}$\gls*{chi}'(\gls*{tau}\gls*{Mem}')\gls*{rho}\gls*{Env}\gls*{phi}\gls*{k}$ \\ \hline

  $\llbracket br\_if\ l \rrbracket_{\gls*{rho}\gls*{chi}\gls*{Mem}\gls*{Env}\gls*{phi}\gls*{k}}\ = 
    let\ \gls*{tau},\gls*{chi}'\ = \gls*{rho}\ l\ in$ \\
  \phantom{XX}$let\ \gls*{Mem}',t\ =\ intbool\ \gls*{Mem}\ in$ & Conditional branching.\\
  \phantom{XX}$t\ ?\ 
    \gls*{chi}'(\gls*{tau}\gls*{Mem}')\gls*{rho}\gls*{Env}\gls*{phi}\gls*{k}\ :\ 
    \gls*{chi}\gls*{Mem}'\gls*{rho}\gls*{Env}\gls*{phi}\gls*{k}$ \\ \hline

  $\llbracket if\ then\ b_{t}\ (c_{t})\ else\ b_{f}\ (c_{f}) \rrbracket_{\gls*{rho}\gls*{chi}\gls*{Mem}\gls*{Env}\gls*{phi}\gls*{k}} =$ & Evaluation of if-then-else.\\
  \phantom{XX}$let\ \gls*{Mem}',t\ =\ intbool\ \gls*{Mem}\ in$ & $b_{t}$, $b_{f}$ are Block labels. \\
  \phantom{XX}$t\ ?\ 
  \llbracket label\ b_{t}\ (c_{t}) 
    \rrbracket_{\gls*{rho}\gls*{chi}\gls*{Mem}'\gls*{Env}\gls*{phi}\gls*{k}}\ : 
  \llbracket label\ b_{f}\ (c_{f}) 
    \rrbracket_{\gls*{rho}\gls*{chi}\gls*{Mem}'\gls*{Env}\gls*{phi}\gls*{k}}$ & \\ \hline 

  \pagebreak

  $\llbracket fun\ (\tilde{p})\ (\tilde{r})\ (c) \rrbracket_
    {\gls*{rho}\gls*{chi}\gls*{Mem}\gls*{Env}\gls*{phi}\gls*{k}} =\ $\\
  \phantom{XX}$let\ \gls*{phi}'\ =\ tick\ \gls*{phi}\ in$ & Evaluation of a function definition:\\
  \phantom{XX}$let\ \gls*{Env}',newvars\ =\ new\gls*{Env}\ \gls*{Env}\ \gls*{phi}'\ (decl\ \tilde{p})\ in$ & as a definition can be recursive, \\
  \phantom{XX}$let\ \gls*{Mem}'\ =\ bind\ \gls*{Mem}\ \gls*{phi}'\ (val\ \Tilde{p})\ newvars\ in$ & we use the $fix$ operator.\\
  \phantom{XX}$let\ \gls*{tau}\ =\ fun\gls*{tau}\ \tilde{r}\ in $\\
  \phantom{XX}$let\ \gls*{k}'\ =\ push\ \gls*{k}\ (\gls*{chi},\gls*{Env},\gls*{tau})\ in$\\
  \phantom{XX}$fix(\lambda\Theta.\llbracket c \rrbracket_{\gls*{rho}\gls*{chi}(\Theta \rightarrow \gls*{Mem}')\gls*{Env}'\gls*{phi}'\gls*{k}'})$\\ \hline 
  
  $\llbracket call\ f_{i} \rrbracket_{\gls*{rho}\gls*{chi}\gls*{Mem}\gls*{Env}\gls*{phi}\gls*{k}} =\ $\\
  \phantom{XX}$let\ f_{body}\ =\ body\ \gls*{mod}.funcs[i]\ in$ & Evaluation of a direct function call:\\
  \phantom{XX}$let\ \gls*{Mem}',\Tilde{v}\ =\ pop\_n\ \gls*{Mem}.sk\ (arity\ \gls*{mod}.funcs[i].t)\ in$ & the body is retrieved from $M$, \\
  \phantom{XX}$let\ \widetilde{params},\Tilde{r}\ = {mod}.funcs[i].params,$ & the function parameters are bound,\\
  \phantom{XXXX}${mod}.funcs[i].return\ in$ & then the body is evaluated.\\
  \phantom{XX}$let\ \tilde{p}\ =\ map\ (\lambda fp,v.(fp,v))\ \widetilde{params},\tilde{v}\ in$\\
  \phantom{XX}$\llbracket fun\ (\tilde{p})\ (\Tilde{r}) \ (f_{body}) \rrbracket_{\gls*{rho}\gls*{chi}\gls*{Mem}'\gls*{Env}\gls*{phi}\gls*{k}}$ \\ \hline
  
  $\llbracket call\_indirect\ t_{i} \rrbracket_{\gls*{rho}\gls*{chi}\gls*{Mem}\gls*{Env}\gls*{phi}\gls*{k}} =\ $ & Evaluation of function indirection.\\
  \phantom{XX}$let\ \gls*{Mem}',f_{i}\ =\ pop\ \gls*{Mem}\ in$ & The reference is held on a table,\\
  \phantom{XX}$let\ f_{idx}\ i\ =\ \gls*{Mem}.se.funtable[i]\ in$ &and it is retrieved by \\
  \phantom{XX}$\llbracket call\ (f_{idx}\ i)\rrbracket_{\gls*{rho}\gls*{chi}\gls*{Mem}'\gls*{Env}\gls*{phi}\gls*{k}}$ & the function $f_{idx}\ i$.\\ \hline

  $\llbracket return \rrbracket_{\gls*{rho}\gls*{chi}\gls*{Mem}\gls*{Env}\gls*{phi}\gls*{k}} =$ \\ 
  \phantom{XX}$let\ \gls*{k}',\gls*{chi}',\gls*{Env}',\gls*{tau} =\ pop\ \gls*{k}\ in$ & Evaluation of return from function.\\
  \phantom{XX}$\gls*{chi}'(\gls*{tau}\gls*{Mem})\gls*{rho}\gls*{Env}\gls*{phi}\gls*{k}'$ \\ \hline

\caption{Concrete semantics for the subset of the WebAssembly language we studied.}
\label{tab:concretesem}
\end{longtable}

\subsubsection{Abstract semantics}
In Table \ref{tab:abstractsem} we introduce the abstract semantics functions. As already mentioned, we only introduce the 
semantics of instructions that are non-homomorphic to their concrete counterparts.
The descriptions on the side provide an explanation of how the abstract semantic functions differ from their concrete version.

\begin{longtable}{l | c}
Abstract semantics & Brief description \\
\hline
\endfirsthead
Abstract semantics & Brief description \\
\hline
\endhead
    $\llbracket br\ l \rrbracket_{\gls*{rho}\gls*{chi}\gls*{aMem}\gls*{aEnv}\gls*{phi}\gls*{aK}}\ =$\\
    \phantom{XX}$let\ \gls*{tau},\gls*{chi}'\ = \gls*{rho}\ l\ in$ & Abstract $br$, with respect to its\\
    \phantom{XX}$let\ \gls*{aMem}'\ =\ \gls*{tau}\gls*{aMem}\ in$ & concrete counterpart, computes the\\
    \phantom{XX}$let\ \gls*{rho}',\gls*{chi}''\ =\ case\ l\ of:$ & widening (\gls*{widen}) or the upper bound (\gls*{sqcup})\\ 
    \phantom{XXXX}$loop \rightarrow\ let\ r\ = \gls*{chi}'\ \gls*{widen}\ \gls*{aMem}'\ in $ & between the ``old" computation $\gls*{chi}'$,\\ 
    \phantom{XXXXXX}$\gls*{rho} [ l \leftarrow (r),\gls*{tau} ],r$ &  and the ``new" one, $\gls*{aMem}'$.\\
    \phantom{XXXX}$block \rightarrow\ let\ r\ =\  \gls*{chi}'\ \gls*{sqcup}\ \gls*{aMem}'\ in$ & With this mechanism we obtain $\gls*{chi}''$.\\ 
    \phantom{XXXXXX}$\gls*{rho} [ l \leftarrow (r),\gls*{tau} ],r$ & The \gls*{widen} operator ensures termination.\\
    \phantom{XX}$in\ \gls*{chi}''\gls*{aMem}'\gls*{rho}'\gls*{aEnv}\gls*{phi}\gls*{aK}$\\ \hline
\pagebreak
    $\llbracket br\_if\ l \rrbracket_{\gls*{rho}\gls*{chi}\gls*{aMem}\gls*{aEnv}\gls*{phi}\gls*{aK}}\ =$\\
    \phantom{XX}$let\ \gls*{tau},\gls*{chi}'\ = \gls*{rho}\ l\ in$ & A function $filter$ is applied.\\
    \phantom{XX}$let\ \gls*{aMem}',\widehat{e}\ =\ \widehat{intbool}\ \gls*{aMem}\ in$ & This improves the precision\\
    \phantom{XX}$let\ \gls*{aMem}'_{t}\ =\ \gls*{tau}(filter\ \widehat{e}\ \gls*{aMem}')\ in$ & of the analysis.\\
    \phantom{XX}$let\ \gls*{aMem}'_{f}\ =\ filter\ \neg\widehat{e}\ \gls*{aMem}'\ in$ & $\gls*{aMem}'_{t}$, $\gls*{aMem}'_{f}$ are the memories in which, \\
    \phantom{XX}$let\ \gls*{rho}',\gls*{chi}''\ =\ case\ l\ of:$ & respectively, $\widehat{e}$ is true and false.\\ 
    \phantom{XXXX}$loop \rightarrow\ let\ r\ =\ \gls*{chi}'\ \gls*{widen}\ \gls*{aMem}'_{t}\ in\ \gls*{rho}$ 
    $[ l \leftarrow (r),\gls*{tau} ],r$\\
    \phantom{XXXX}$block \rightarrow\ let\ r\ =\  \gls*{chi}'\ \gls*{sqcup}\ \gls*{aMem}'_{t}\ in\ \gls*{rho}$ 
    $[ l \leftarrow (r),\gls*{tau} ],r$\\
    \phantom{XX}$in\ \gls*{chi}''\gls*{aMem}'_{t}\gls*{rho}'\gls*{aEnv}\gls*{phi}\gls*{aK}\ \gls*{sqcup}\ $ 
    $\gls*{chi}\gls*{aMem}'_{f}\gls*{rho}\gls*{aEnv}\gls*{phi}\gls*{aK}$  \\ \hline
      
    $\llbracket 
      if\ then\ b_{t}\ (c_{t};end)\ else\ b_{f}\ (c_{f};end) 
      \rrbracket_{\gls*{rho}\gls*{chi}\gls*{aMem}\gls*{aEnv}\gls*{phi}\gls*{aK}} =$ & Computes the upper bound between\\
    \phantom{XX}$let\ \gls*{aMem}',\widehat{e}\ =\ \widehat{intbool}\ \gls*{aMem}\ in$ & the computations in $b_{t}$ and $b_{f}$.\\
    \phantom{XX}$\llbracket label\ b_{t}\ (c_{t};end) 
      \rrbracket_{\gls*{rho}\gls*{chi}(filter\ \widehat{e}\ \gls*{aMem}')\gls*{aEnv}\gls*{phi}\gls*{aK}}\ 
      \gls*{sqcup}$ & Similarly to $br\_if$, $filter$ is applied.\\  
    \phantom{XXXX}$\llbracket label\ b_{f}\ (c_{f};end)
      \rrbracket_{\gls*{rho}\gls*{chi}(filter\ \neg \widehat{e}\ \gls*{aMem}')\gls*{aEnv}\gls*{phi}\gls*{aK}}$ \\ \hline
      
    $\llbracket call\_indirect\ t_{i} \rrbracket_{\gls*{rho}\gls*{chi}\gls*{aMem}\gls*{aEnv}\gls*{phi}\gls*{aK}} =\ $ & Computes the upper bound between\\
    \phantom{XX}$let\ \gls*{aMem}',\widehat{f_{i}}\ =\ pop\ \gls*{aMem}\ in$ & the calls to the possible callees. \\
    \phantom{XX}$let\ f_{idx}\ i\ =\ \gls*{aMem}.se.funtable[i]\ in$ & Target functions are type filtered. \\
    \phantom{XX}$\gls*{sqcup}(\{\forall\ i\ \in \gls*{gamma}(\widehat{f_{i}})\}\mid
    \llbracket call\ (f_{idx}\ i)\rrbracket_{\gls*{rho}\gls*{chi}\gls*{aMem}'\gls*{aEnv}\gls*{phi}\gls*{aK}}\})$ & The notation omits this detail for\\
    & triviality and readability.\\

\caption{Abstract semantics for the subset of the WebAssembly language we studied.}
\label{tab:abstractsem}
\end{longtable}

In this semantic notation, we omit the specific semantics of the call graph for readability.
Briefly explained: any instruction that is not \verb|call i| or \verb|call_indirect t| will produce an 
empty set of callees $\phi$. The \verb|call i| instruction will produce a singleton \verb|{i}|, and
the \verb|call_indirect| instruction will produce a set containing the function indexes obtained by resolving the indirections generated by concretizing the 
abstract value on top of the stack. As noted, we also filter possible callee functions that do not match the type 
\verb|t|.

\subsection{Architecture of \tool}
As mentioned, \tool is a specialized tool for constructing call graphs in a ``reasonably" precise, 
yet sound, manner. We can achieve this by deducing what value(s) can be on top of the stack when a 
\verb+call_indirect+ instruction is met. This can be done with a varying degree of precision. 
For our work, we use a relational abstraction to represent values. We then build specialized data 
structures on top of it to represent the operand stack, the call stack, and the linear memory. 
As \verb|memory| operations are outside the scope of this work, and the call stack does not present 
any particular novelty, we will not discuss those and will focus solely on the operand stack.

One of the peculiarities of our implementation is its capability to analyze WebAssembly code directly as 
stack machines, relationally, and lazily. 
This is not the case for other analyzers for languages based on stack machines. For example, 
\emph{Sturdy}~\cite{sturdy_combinatorfix,sturdy_wasm_ai2023}, to the best of our knowledge, 
is only capable of non-relational analyses when considering stack machines. 
Other tools, such as \emph{Soot}~\cite{soot}, \emph{Wimpl}~\cite{wimpl}, and 
\emph{cmm\_of\_wasm}\footnote{\url{https://github.com/SimonJF/cmm\_of\_wasm}} employ some kind of 
explicit language transformation step.

The second peculiarity of our implementation of the operand stack consists of lazily building 
abstract expressions on the fly, without necessarily evaluating them (i.e. if we are adding two 
integer operands together and we are not interested in integer overflows/underflows, we do not 
evaluate the expression, we pop \verb|e1|, \verb|e2| from the stack and push the resulting mathematical expression \verb|e1+e2| to the operand stack). 
This applies only to the cases in which it is possible to construct an expression representing the 
content of the stack lazily. For instance, bitwise operations are not performed this way.

\subsection{Discussion}\label{subs:evaluation}
As the tool is under development a real experimental evaluation cannot be performed yet. Ignoring performance issues, we demonstrate how
the tool operates on the common language concerns
in WebAssembly. The systematic literature review in \cite{toughcall} exhaustively showcases all of the language concerns when we are interested in building call graphs.

We will also make some brief considerations about the language concerns outside the scope of 
this paper: mutations performed by the host and \verb|memory| instructions.

The work presented in \cite{toughcall} assesses that the main language concerns that are
computed in an unsound manner are the reachability of functions in imported and exported tables.
In WebAssembly the reachability of references in exported tables should be given for granted,
as they are equivalently reachable as exported functions. It can be helpful to think of them as a 
mutable set of exported function references as opposed to a set of exported functions.

Regarding imported functions, while we do not 
have any information about the behavior of such functions, unless some form of extra information is available, we can always asses that they are reachable
and can give an unconstrained function output of the correct typing, as it is explicit in the 
function signature.

Besides these points and concerns relative to unknown imported objects, we are potentially able to 
cover most of the rest of the concerns with at least the same degree of precision as the state 
of the art, while retaining soundness, as we use inference rules which consider the semantics 
of a given program instead of type inference only, and a relational abstraction that allows to
derive stronger properties.

Some specific language concerns, such as ``10: Table init. offset is imported 
from host" would need some form of specification to provide a conclusive analysis. 
This means that with no additional information available we either produce a sound and very 
imprecise analysis or an unsound one, which is not a desirable result for the scope of this work. 
This need for additional information takes us to the next point: table mutability.

The concern of table mutability performed by the host is, in a fundamental way, similar to the 
previous one: if we are to produce a sound analysis we should assume that calling a host function 
can mutate the tables, and without information available we are ``forced" to be conservative and 
assume we do not know what function reference is held at any given table entry, making the analysis 
of some programs as imprecise as an analysis based on function types only. 

We can generalize these concerns as ``Target callee is dependant on host behavior".
This covers the concerns: ``9: Table is mutated by host", ``10: Table init. offset is imported from 
host", ``11: Memory init. offset is imported from host".

Further indications about this issue will be discussed in Section \ref*{sec:futurework}.
\section{Related work}\label{sec:related}
WebAssembly was first released in 2017~\cite{haas2017bringing} and can therefore be considered a relatively ``young" language. The formal specification and a reference implementation in OCaml are available online~\cite{WebAssemblyCoreSpecification1}. An Isabelle specification is also available in~\cite{watt2018mechanising}.

The work presented in~\cite{255318} shows how WebAssembly can be used as an attack vehicle to write to memory and cause unexpected behavior in the host environment. The tool \emph{Wasmati} introduced in~\cite{wasmati} employs a graph-based representation of WebAssembly code, called Code Property Graphs (CPG), and checks security properties on these using a query language. Wasmati requires a call graph but can generate one by considering function signatures only. 

Even in the absence of buffer overflows and similar attacks, it is still possible for WebAssembly to leak information in unintended ways; to address this issue, recent work has focused on information flow analysis, see for instance~\cite{stievenart2020compositional} and~\cite{iulia22}. 

Tools that focus, more in general, on static analysis of WebAssembly include \emph{WASP}~\cite{WASP}, \emph{Manticore}~\cite{mossberg2019manticore}, and \emph{Wassail}~\cite{stievenart2021wassail}. The first two tools are  based, respectively, on symbolic and concolic execution. Wassail, instead, can be seen as a library to develop analyses. \emph{Wassail} also supports \emph{program slicing}~\cite{wassail_slicing} and the computation of control flow graphs and call graphs. In particular, similar to \emph{Wasmati}, it can provide an over-approximation of the call graph for a specific WebAssembly module by relying on the types of functions invoked in indirect calls  (see Section ~\ref{sec:background}). \emph{Sturdy}~\cite{sturdy_wasm_ai2023,sturdy_combinatorfix} presents another framework to develop and perform static analysis. It is a modular, extensible framework based on abstract interpretation. It supports WebAssembly bytecode and out-of-the-box can perform analyses such as dead code analysis and taint analysis.

The work presented in \cite{wimpl} uses a different approach to verifying WebAssembly code. The authors introduce a simplified C-like Intermediate Representation called \emph{Wimpl}. The goal is to use existing tools for C instead of developing verification tools from scratch.

The work in \cite{toughcall} presents a comparison between tools that are capable of constructing a call graph for WebAssembly dynamically or statically, introducing a set of micro-benchmarks that allow the assessment of the precision of a certain analyzer via specifically crafted edge cases. Such edge cases cover the language concerns regarding indirect calls.
\section{Conclusion}\label{sec:conclusion}
Given the increasing importance of WebAssembly and the importance of having static analysis support,
we have given concrete and abstract semantics for a simplified version of the WebAssembly language. 
We also highlighted some additional challenges which we do not cover in this work and propose 
means to tackle them. 

The abstract semantics we specified form the basis of an analyzer based on the theory 
of Abstract Interpretation that is currently being developed.

We also described the approach we are pursuing in the implementation of the analyzer, which features 
a lazy operand stack and the ability to directly analyze WebAssembly bytecode relationally. 
Based on this, we can predict our approach to be at least as precise as the sound ones present 
in the state-of-the-art, whilst potentially being capable of tackling language concerns which, 
at the moment, are being computed in an unsound manner.
\section{Future work}\label{sec:futurework}
The ability of WebAssembly to work as an external mechanism to provide high-performance
computations is what makes it so popular. Unfortunately, we cannot assume in a definitive manner that 
all WebAssembly modules work in isolation and are closed. Some possible future work 
in improving the analysis of WebAssembly modules can and should include some mechanism for 
defining or inferring host-module(s) interactions. 

This includes analysis of imported host functions (and subsequently analysis of the 
functions depending on them) and production of function summaries or a form of 
specification in a custom Domain Specific Language.

Trivially, a DSL would allow to manually set the rules we would instead infer, and would
work as a sort of specification contract (i.e: the soundness of the analysis is 
guaranteed with respect to the contract). A similar approach has been explored in~\cite{eunomia} 
for symbolic execution but, to the best of our knowledge, no approaches are available for 
abstract interpretation.

Both of these approaches have pros and cons: the former keeps the analysis fully 
automated, but the analysis still has to maintain soundness at the cost of precision. 
The latter requires the handwriting of a specification and a degree of knowledge 
about the specific host program interacting with the WebAssembly module: it is potentially 
more precise, as false positives can be ruled out, but it is prone to be influenced by 
human error.
\label{sect:bib}
\bibliography{easychair}

\begin{thebibliography}{23}
\expandafter\ifx\csname natexlab\endcsname\relax\def\natexlab#1{#1}\fi
\providecommand{\url}[1]{\texttt{#1}}
\providecommand{\href}[2]{#2}
\providecommand{\path}[1]{#1}
\providecommand{\DOIprefix}{doi:}
\providecommand{\ArXivprefix}{arXiv:}
\providecommand{\URLprefix}{URL: }
\providecommand{\Pubmedprefix}{pmid:}
\providecommand{\doi}[1]{\href{http://dx.doi.org/#1}{\path{#1}}}
\providecommand{\Pubmed}[1]{\href{pmid:#1}{\path{#1}}}
\providecommand{\bibinfo}[2]{#2}
\ifx\xfnm\relax \def\xfnm[#1]{\unskip,\space#1}\fi
\bibitem[{Ene(2022)}]{alexene_wasm}
\bibinfo{author}{A.~Ene}, \bibinfo{title}{How prime video updates its app for more than 8,000 device types}, \bibinfo{howpublished}{\url{https://www.amazon.science/blog/how-prime-video-updates-its-app-for-more-than-8-000-device-types}}, \bibinfo{year}{2022}. \bibinfo{note}{Accessed: 2022-10-11}.
\bibitem[{Schroeder et~al.(2021)Schroeder, Jordan, Schulz, Cain, Hanley, and Fay}]{schroeder_disneyplus}
\bibinfo{author}{T.~Schroeder}, \bibinfo{author}{D.~Jordan}, \bibinfo{author}{S.~Schulz}, \bibinfo{author}{R.~Cain}, \bibinfo{author}{M.~Hanley}, \bibinfo{author}{M.~Fay}, \bibinfo{title}{Introducing the disney+ application development kit (adk)}, \bibinfo{howpublished}{\url{https://medium.com/disney-streaming/introducing-the-disney-application-development-kit-adk-ad85ca139073}}, \bibinfo{year}{2021}. \bibinfo{note}{Accessed: 2022-10-10}.
\bibitem[{Lehmann et~al.(2023)Lehmann, Thalakottur, Tip, and Pradel}]{toughcall}
\bibinfo{author}{D.~Lehmann}, \bibinfo{author}{M.~Thalakottur}, \bibinfo{author}{F.~Tip}, \bibinfo{author}{M.~Pradel},
\newblock \bibinfo{title}{That's a tough call: Studying the challenges of call graph construction for webassembly},
\newblock in: \bibinfo{editor}{R.~Just}, \bibinfo{editor}{G.~Fraser} (Eds.), \bibinfo{booktitle}{Proceedings of the 32nd {ACM} {SIGSOFT} International Symposium on Software Testing and Analysis, {ISSTA} 2023, Seattle, WA, USA, July 17-21, 2023}, \bibinfo{publisher}{{ACM}}, \bibinfo{year}{2023}, pp. \bibinfo{pages}{892--903}. \URLprefix \url{https://doi.org/10.1145/3597926.3598104}. \DOIprefix\doi{10.1145/3597926.3598104}.
\bibitem[{Sti{\'e}venart and De~Roover(2021)}]{stievenart2021wassail}
\bibinfo{author}{Q.~Sti{\'e}venart}, \bibinfo{author}{C.~De~Roover},
\newblock \bibinfo{title}{Wassail: a webassembly static analysis library},
\newblock in: \bibinfo{booktitle}{Fifth International Workshop on Programming Technology for the Future Web}, \bibinfo{year}{2021}.
\bibitem[{Brandl et~al.(2023)Brandl, Erdweg, Keidel, and Hansen}]{sturdy_wasm_ai2023}
\bibinfo{author}{K.~Brandl}, \bibinfo{author}{S.~Erdweg}, \bibinfo{author}{S.~Keidel}, \bibinfo{author}{N.~Hansen},
\newblock \bibinfo{title}{Modular abstract definitional interpreters for webassembly},
\newblock in: \bibinfo{editor}{K.~Ali}, \bibinfo{editor}{G.~Salvaneschi} (Eds.), \bibinfo{booktitle}{37th European Conference on Object-Oriented Programming, {ECOOP} 2023, July 17-21, 2023, Seattle, Washington, United States}, volume \bibinfo{volume}{263} of \textit{\bibinfo{series}{LIPIcs}}, \bibinfo{publisher}{Schloss Dagstuhl - Leibniz-Zentrum f{\"{u}}r Informatik}, \bibinfo{year}{2023}, pp. \bibinfo{pages}{5:1--5:28}. \URLprefix \url{https://doi.org/10.4230/LIPIcs.ECOOP.2023.5}. \DOIprefix\doi{10.4230/LIPIcs.ECOOP.2023.5}.
\bibitem[{Mossberg et~al.(2019)Mossberg, Manzano, Hennenfent, Groce, Grieco, Feist, Brunson, and Dinaburg}]{mossberg2019manticore}
\bibinfo{author}{M.~Mossberg}, \bibinfo{author}{F.~Manzano}, \bibinfo{author}{E.~Hennenfent}, \bibinfo{author}{A.~Groce}, \bibinfo{author}{G.~Grieco}, \bibinfo{author}{J.~Feist}, \bibinfo{author}{T.~Brunson}, \bibinfo{author}{A.~Dinaburg},
\newblock \bibinfo{title}{Manticore: {A} user-friendly symbolic execution framework for binaries and smart contracts},
\newblock in: \bibinfo{booktitle}{34th {IEEE/ACM} International Conference on Automated Software Engineering, {ASE} 2019, San Diego, CA, USA, November 11-15, 2019}, \bibinfo{publisher}{{IEEE}}, \bibinfo{year}{2019}, pp. \bibinfo{pages}{1186--1189}. \URLprefix \url{https://doi.org/10.1109/ASE.2019.00133}. \DOIprefix\doi{10.1109/ASE.2019.00133}.
\bibitem[{Sti{\'e}venart and De~Roover(2020)}]{stievenart2020compositional}
\bibinfo{author}{Q.~Sti{\'e}venart}, \bibinfo{author}{C.~De~Roover},
\newblock \bibinfo{title}{Compositional information flow analysis for webassembly programs},
\newblock in: \bibinfo{booktitle}{2020 IEEE 20th International Working Conference on Source Code Analysis and Manipulation (SCAM)}, \bibinfo{organization}{IEEE}, \bibinfo{year}{2020}, pp. \bibinfo{pages}{13--24}.
\bibitem[{Nguyen et~al.(2013)Nguyen, Nguyen, and Quan}]{nguyen2013binarycfg}
\bibinfo{author}{M.~H. Nguyen}, \bibinfo{author}{T.~B. Nguyen}, \bibinfo{author}{T.~T. Quan},
\newblock \bibinfo{title}{A hybrid aproach for control flow graph construction from binary code},
\newblock in: \bibinfo{booktitle}{20th Asia-Pacific Software Engineering Conference}, \bibinfo{year}{2013}.
\bibitem[{Flores{-}Montoya and Schulte(2020)}]{montoya2020ddisasm}
\bibinfo{author}{A.~Flores{-}Montoya}, \bibinfo{author}{E.~M. Schulte},
\newblock \bibinfo{title}{Datalog disassembly},
\newblock in: \bibinfo{editor}{S.~Capkun}, \bibinfo{editor}{F.~Roesner} (Eds.), \bibinfo{booktitle}{29th {USENIX} Security Symposium, {USENIX} Security 2020, August 12-14, 2020}, \bibinfo{publisher}{{USENIX} Association}, \bibinfo{year}{2020}, pp. \bibinfo{pages}{1075--1092}. \URLprefix \url{https://www.usenix.org/conference/usenixsecurity20/presentation/flores-montoya}.
\bibitem[{Dean et~al.(1995)Dean, Grove, and Chambers}]{dean1995optimization}
\bibinfo{author}{J.~Dean}, \bibinfo{author}{D.~Grove}, \bibinfo{author}{C.~Chambers},
\newblock \bibinfo{title}{Optimization of object-oriented programs using static class hierarchy analysis},
\newblock in: \bibinfo{booktitle}{European Conference on Object-Oriented Programming}, \bibinfo{organization}{Springer}, \bibinfo{year}{1995}, pp. \bibinfo{pages}{77--101}.
\bibitem[{Brito et~al.(2022)Brito, Lopes, Santos, and Santos}]{wasmati}
\bibinfo{author}{T.~Brito}, \bibinfo{author}{P.~Lopes}, \bibinfo{author}{N.~Santos}, \bibinfo{author}{J.~F. Santos},
\newblock \bibinfo{title}{Wasmati: An efficient static vulnerability scanner for webassembly},
\newblock \bibinfo{journal}{Computers \& Security} \bibinfo{volume}{118} (\bibinfo{year}{2022}) \bibinfo{pages}{102745}. \URLprefix \url{https://www.sciencedirect.com/science/article/pii/S0167404822001407}. \DOIprefix\doi{https://doi.org/10.1016/j.cose.2022.102745}.
\bibitem[{Cousot and Cousot(1977)}]{cousot_ai}
\bibinfo{author}{P.~Cousot}, \bibinfo{author}{R.~Cousot},
\newblock \bibinfo{title}{Abstract interpretation: {A} unified lattice model for static analysis of programs by construction or approximation of fixpoints},
\newblock in: \bibinfo{editor}{R.~M. Graham}, \bibinfo{editor}{M.~A. Harrison}, \bibinfo{editor}{R.~Sethi} (Eds.), \bibinfo{booktitle}{Conference Record of the Fourth {ACM} Symposium on Principles of Programming Languages, Los Angeles, California, USA, January 1977}, \bibinfo{publisher}{{ACM}}, \bibinfo{year}{1977}, pp. \bibinfo{pages}{238--252}. \URLprefix \url{https://doi.org/10.1145/512950.512973}. \DOIprefix\doi{10.1145/512950.512973}.
\bibitem[{Keidel et~al.(2023)Keidel, Erdweg, and Homb{\"{u}}cher}]{sturdy_combinatorfix}
\bibinfo{author}{S.~Keidel}, \bibinfo{author}{S.~Erdweg}, \bibinfo{author}{T.~Homb{\"{u}}cher},
\newblock \bibinfo{title}{Combinator-based fixpoint algorithms for big-step abstract interpreters},
\newblock \bibinfo{journal}{Proc. {ACM} Program. Lang.} \bibinfo{volume}{7} (\bibinfo{year}{2023}) \bibinfo{pages}{955--981}. \URLprefix \url{https://doi.org/10.1145/3607863}. \DOIprefix\doi{10.1145/3607863}.
\bibitem[{Vall{\'{e}}e{-}Rai et~al.(1999)Vall{\'{e}}e{-}Rai, Co, Gagnon, Hendren, Lam, and Sundaresan}]{soot}
\bibinfo{author}{R.~Vall{\'{e}}e{-}Rai}, \bibinfo{author}{P.~Co}, \bibinfo{author}{E.~Gagnon}, \bibinfo{author}{L.~J. Hendren}, \bibinfo{author}{P.~Lam}, \bibinfo{author}{V.~Sundaresan},
\newblock \bibinfo{title}{Soot - a java bytecode optimization framework},
\newblock in: \bibinfo{editor}{S.~A. MacKay}, \bibinfo{editor}{J.~H. Johnson} (Eds.), \bibinfo{booktitle}{Proceedings of the 1999 conference of the Centre for Advanced Studies on Collaborative Research, November 8-11, 1999, Mississauga, Ontario, Canada}, \bibinfo{publisher}{{IBM}}, \bibinfo{year}{1999}, p.~\bibinfo{pages}{13}. \URLprefix \url{https://dl.acm.org/citation.cfm?id=782008}.
\bibitem[{Thalakottur et~al.(2022)Thalakottur, Tip, Lehmann, and Pradel}]{wimpl}
\bibinfo{author}{M.~Thalakottur}, \bibinfo{author}{F.~Tip}, \bibinfo{author}{D.~Lehmann}, \bibinfo{author}{M.~Pradel},
\newblock \bibinfo{title}{Wimpl: A simple ir for static analysis of webassembly binaries},
\newblock in: \bibinfo{booktitle}{Program Analysis for WebAssembly (PAW) 2022}, \bibinfo{year}{2022}.
\bibitem[{Haas et~al.(2017)Haas, Rossberg, Schuff, Titzer, Holman, Gohman, Wagner, Zakai, and Bastien}]{haas2017bringing}
\bibinfo{author}{A.~Haas}, \bibinfo{author}{A.~Rossberg}, \bibinfo{author}{D.~L. Schuff}, \bibinfo{author}{B.~L. Titzer}, \bibinfo{author}{M.~Holman}, \bibinfo{author}{D.~Gohman}, \bibinfo{author}{L.~Wagner}, \bibinfo{author}{A.~Zakai}, \bibinfo{author}{J.~Bastien},
\newblock \bibinfo{title}{Bringing the web up to speed with webassembly},
\newblock in: \bibinfo{booktitle}{Proceedings of the 38th ACM SIGPLAN Conference on Programming Language Design and Implementation (PLDI)}, \bibinfo{year}{2017}, pp. \bibinfo{pages}{185--200}.
\bibitem[{Rossberg(2019)}]{WebAssemblyCoreSpecification1}
\bibinfo{author}{A.~Rossberg}, \bibinfo{title}{{WebAssembly Core Specification}}, \bibinfo{year}{2019}. \URLprefix \url{https://www.w3.org/TR/wasm-core-1/}.
\bibitem[{Watt(2018)}]{watt2018mechanising}
\bibinfo{author}{C.~Watt},
\newblock \bibinfo{title}{Mechanising and verifying the webassembly specification},
\newblock in: \bibinfo{booktitle}{Proceedings of the 7th ACM SIGPLAN International Conference on certified programs and proofs}, \bibinfo{year}{2018}, pp. \bibinfo{pages}{53--65}.
\bibitem[{Lehmann et~al.(2020)Lehmann, Kinder, and Pradel}]{255318}
\bibinfo{author}{D.~Lehmann}, \bibinfo{author}{J.~Kinder}, \bibinfo{author}{M.~Pradel},
\newblock \bibinfo{title}{Everything old is new again: Binary security of {WebAssembly}},
\newblock in: \bibinfo{booktitle}{29th USENIX Security Symposium (USENIX Security 20)}, \bibinfo{publisher}{USENIX Association}, \bibinfo{year}{2020}, pp. \bibinfo{pages}{217--234}. \URLprefix \url{https://www.usenix.org/conference/usenixsecurity20/presentation/lehmann}.
\bibitem[{Bastys et~al.(2022)Bastys, Algehed, Sj{\"o}sten, and Sabelfeld}]{iulia22}
\bibinfo{author}{I.~Bastys}, \bibinfo{author}{M.~Algehed}, \bibinfo{author}{A.~Sj{\"o}sten}, \bibinfo{author}{A.~Sabelfeld},
\newblock \bibinfo{title}{Secwasm: Information flow control for webassembly},
\newblock in: \bibinfo{booktitle}{Program Analysis for WebAssembly (PAW)}, \bibinfo{year}{2022}.
\bibitem[{Marques et~al.(2022)Marques, Fragoso~Santos, Santos, and Ad\~{a}o}]{WASP}
\bibinfo{author}{F.~Marques}, \bibinfo{author}{J.~Fragoso~Santos}, \bibinfo{author}{N.~Santos}, \bibinfo{author}{P.~Ad\~{a}o},
\newblock \bibinfo{title}{{Concolic Execution for WebAssembly}},
\newblock in: \bibinfo{editor}{K.~Ali}, \bibinfo{editor}{J.~Vitek} (Eds.), \bibinfo{booktitle}{36th European Conference on Object-Oriented Programming (ECOOP 2022)}, volume \bibinfo{volume}{222} of \textit{\bibinfo{series}{Leibniz International Proceedings in Informatics (LIPIcs)}}, \bibinfo{publisher}{Schloss Dagstuhl -- Leibniz-Zentrum f{\"u}r Informatik}, \bibinfo{address}{Dagstuhl, Germany}, \bibinfo{year}{2022}, pp. \bibinfo{pages}{11:1--11:29}. \URLprefix \url{https://drops.dagstuhl.de/opus/volltexte/2022/16239}. \DOIprefix\doi{10.4230/LIPIcs.ECOOP.2022.11}.
\bibitem[{Sti{\'e}venart et~al.(2022)Sti{\'e}venart, Binkley, and De~Roover}]{wassail_slicing}
\bibinfo{author}{Q.~Sti{\'e}venart}, \bibinfo{author}{D.~W. Binkley}, \bibinfo{author}{C.~De~Roover},
\newblock \bibinfo{title}{Static stack-preserving intra-procedural slicing of webassembly binaries},
\newblock in: \bibinfo{booktitle}{Proceedings of the 44th International Conference on Software Engineering}, \bibinfo{year}{2022}, pp. \bibinfo{pages}{2031--2042}.
\bibitem[{He et~al.(2023)He, Zhao, Wang, Hu, Guo, Wang, Liang, Li, Chen, and Guo}]{eunomia}
\bibinfo{author}{N.~He}, \bibinfo{author}{Z.~Zhao}, \bibinfo{author}{J.~Wang}, \bibinfo{author}{Y.~Hu}, \bibinfo{author}{S.~Guo}, \bibinfo{author}{H.~Wang}, \bibinfo{author}{G.~Liang}, \bibinfo{author}{D.~Li}, \bibinfo{author}{X.~Chen}, \bibinfo{author}{Y.~Guo},
\newblock \bibinfo{title}{Eunomia: Enabling user-specified fine-grained search in symbolically executing webassembly binaries},
\newblock in: \bibinfo{editor}{R.~Just}, \bibinfo{editor}{G.~Fraser} (Eds.), \bibinfo{booktitle}{Proceedings of the 32nd {ACM} {SIGSOFT} International Symposium on Software Testing and Analysis, {ISSTA} 2023, Seattle, WA, USA, July 17-21, 2023}, \bibinfo{publisher}{{ACM}}, \bibinfo{year}{2023}, pp. \bibinfo{pages}{385--397}. \URLprefix \url{https://doi.org/10.1145/3597926.3598064}. \DOIprefix\doi{10.1145/3597926.3598064}.

\end{thebibliography}

\pagebreak
\appendix
\printunsrtglossary[type=symbols,style=long]
\end{document}